\documentstyle[aps,preprint,epsfig]{revtex}
\tightenlines
\newcommand{\r} {{\mathbf r}}
\newcommand{\si} {\sigma i}
\newcommand{\sj} {\sigma j}
\newcommand{\sji} {\sigma ji}
\newcommand{\ns} {n_\sigma}
\newcommand{\nsi} {n_{\sigma i}}
\newcommand{\nsj} {n_{\sigma j}}
\newcommand{\Exc} {E_{xc}}
\newcommand{\Ex} {E_{x}}
\newcommand{\Vxc} {V_{xc}}
\newcommand{\Vxcs} {V_{xc,\sigma}}

\newcommand{\uxcsi} {u_{xc,\sigma i}}

\newcommand{\Vx} {V_{x}}
\newcommand{\Vxs} {V_{x,\sigma}}

\newcommand{\VBxsi} {\bar{V}_{x,\sigma i}}
\newcommand{\uxsi} {u_{x,\sigma i}}
\newcommand{\uBxsi} {\bar{u}_{x,\sigma i}}
\newcommand{\fsi} {f_{\sigma i}}
\newcommand{\fsj} {f_{\sigma j}}
\newcommand{\psisi} {\psi_{\sigma i}}
\newcommand{\psisj} {\psi_{\sigma j}}
\newcommand{\Ns} {N_{\sigma}}

\begin{document}
\draft

%% ------------------------------------------------------------------- %%

\title{Density-functional Study of Small Molecules within the
	Krieger-Li-Iafrate Approximation}

\author{Yong-Hoon Kim,$^{(a)}$
	Martin St\"adele,$^{(a,b)}$ and
	Richard M. Martin$^{(a,c)}$}
\address{Department of Physics,$^{(a)}$
	Beckman Institute for Advanced Science and Technology,$^{(b)}$
	and Materials Research Laboratory,$^{(c)}$
	University of Illinois at Urbana-Champaign,
	Urbana, Illinois 61801}

\date{ \today }

\maketitle
\begin{abstract}
We report density-functional studies of several small molecules
($H_{2}, N_{2}, CO, H_{2}O$, and $CH_{4}$) within the
Krieger-Li-Iafrate (KLI) approximation to the exact Kohn-Sham
local exchange potential, using a three-dimensional real-space
finite-difference pseudopotential method. It is found that
exchange-only KLI leads to markedly improved eigenvalue spectra
compared to those obtained within the standard local-density
approximation (LDA), the generalized gradient approximation (GGA),
and the Hartree-Fock (HF) method. For structural properties,
exchange-only KLI results are close to the corresponding HF
values. We find that the addition of LDA or GGA correlation energy
functionals to the KLI exact exchange energy functional does not
lead to systematic improvements.
\end{abstract}


\narrowtext

%% ------------------------------------------------------------------- %%
\section{Introduction}
\label{sec:intro}

Density functional theory (DFT) within the Kohn-Sham (KS)
single-particle formalism has now been established as a standard
tool for the study of inhomogeneous quantum many-electron systems.
\cite{HK64,KS65,Par89} The local density approximation (LDA) and
generalized gradient approximation (GGA) are successful in
describing the ground state properties of various systems. On the
other hand, their eigenvalue spectra are known to be far from the
reality, and it has been assumed that KS orbital eigenvalues are
auxiliary quantities without any physical meaning except the
highest occupied one.\cite{Par89} However, it is clear that DFT is
also a legitimate route to the study of excited state properties
via perturbation theory\cite{Gor95} or time-dependent DFT.
\cite{Gro96,Cas96} Especially, G\"orling has recently formulated a
KS formalism for excited states based on DFT perturbation theory
and demonstrated that the KS eigenvalue difference is a
well-defined zeroth order approximation to excitation
energies.\cite{Gor96} This has been also explicitly confirmed for
several atoms by accurate quantum Monte-Carlo (QMC)
calculations.\cite{Umr98,Fil97}

It appears that - at least for localized systems - the unrealistic
description of eigenvalue spectra is mainly due to the
deficiencies of the conventional approximations for the
exchange-correlation energy functional such as the LDA or GGA
rather than the KS scheme itself. The origin of these deficiencies
is now more or less understood from the investigations of the
exact properties of the KS exchange-correlation potential
$\Vxc^{KS}$ which the LDA or GGA does not satisfy: (i)
$\Vxc^{KS}(\r) \rightarrow -1/r$ as $r \rightarrow \infty$ for
finite systems,\cite{Alm85,Per84} (ii) the eigenvalue of the
highest occupied single particle states is the negative of the
ionization energy,\cite{Alm85,Per82} (iii) $\Vxc^{KS}(\r)$ is
self-interaction free, and (iv) $\Vxc^{KS}(\r)$ exhibits an
integer discontinuity upon the addition of an electron.
\cite{Per83,Sha83} However, it has been recognized that
approximations to the {\it local} KS potential obtained from the
{\it orbital} dependent exact exchange energy functional through
the so-called optimized effective potential (OEP)
method\cite{Sha53,Tal76} closely observe these
conditions.\cite{Nor84,Kri90}

In spite of these good features, the implementation of the OEP
method is very complicated and its computational cost is
excessively high. Only spherical atoms,\cite{Tal76,Nor84,Eng93}
jellium surfaces,\cite{Kro85} jellium spheres,\cite{Bul95} and
periodic solids with a small unit cell\cite{Sta97} have been
extensively studied, and there exists no investigation of
molecular systems with the OEP method. However, about a decade
ago, Krieger, Li, and Iafrate (KLI) proposed an approximation
scheme to the OEP, and have shown that the KLI potential with the
exact exchange-only energy functional (KLI-x) satisfies the above
mentioned properties (i)-(iv) of $\Vxc^{KS}$, and accurately
reproduces the OEP-x results for spherical atoms.\cite{Kri90}
Within the KLI-x approximation, the workload in constructing the
potential is significantly reduced, but this method is still
relatively complicated compared with the LDA or GGA and studies
are therefore limited. Atomic systems have been originally
investigated by KLI,\cite{Kri90} and more recently by Grabo {\it
et al.}\cite{Gra98} Bylander and Kleinman studied semiconductors
with the KLI potential.\cite{Byl95} Although molecules are a very
good testing ground of any functional approximation, only some
diatomic molecules have been investigated by Grabo {\it et
al.}\cite{Gra98}

In this article, we report a study of several closed shell
diatomic and polyatomic molecules within the KLI scheme. The
questions we address are the performance of the KLI approximation
to the OEP with the exact exchange-only functional (KLI-x) and
supplemented by LDA and Perdew-Burke-Ernzerhof GGA\cite{Per96}
correlation energy functionals (KLI-x$+$LDA-c and KLI-x$+$GGA-c)
for both structural properties and eigenvalue spectra. The
comparison with and qualitative differences from the LDA/GGA and
the Hartree-Fock (HF) method will be emphasized. In addition,
since we incorporate the exact exchange, we can examine the
accuracy of the LDA and GGA correlation energy functionals. The
underlying computational algorithm is the three-dimensional (3D)
finite-difference pseudopotential method\cite{Che94,YHK2} and, to
our knowledge, this is the first 3D KLI investigation of molecular
systems.

The organization of the paper is as follows. In Sec.
\ref{sec:formalism}, the formulation of KLI is reviewed and the
differences of the KLI-x from the HF method will be emphasized.
The implementation of KLI-x into our molecular code and its test
results are described. In Sec. \ref{sec:results}, results and
discussions for several molecules will be presented for both
structural properties and eigenvalue spectra. Especially,
comparisons between the KLI-x and the LDA/GGA or HF method are
made, and physical origins of these differences will be discussed.
In Sec. \ref{sec:conclusions}, we conclude the paper by
summarizing our work and future prospects.

%% ------------------------------------------------------------------- %%
\section{Formalism}
\label{sec:formalism}

\subsection{Approximation of OEP by Krieger, Li, and Iafrate}
\label{subsec:KLI}

We start from the spin-dependent KS equations for
KS orbitals $\psi_{\si}(\r)$ for spin channel $\sigma = \uparrow, \downarrow$
and KS eigenvalues $\epsilon_{\si}$,
with the possibility of fractional occupation numbers $\fsi$
(Hartree atomic units are used throughout the paper)
\begin{equation}
\label{eq:KS}
\biggl[ -\frac{1}{2} \nabla^2 + V^{KS}_{\sigma}(\r) \biggr] \psisi (\r) =
\epsilon_{\si} \psi_{\si}(\r),
\end{equation}
with the spin density
\begin{equation}
\label{eq:n1}
\ns(\r) = \sum_{\i=1}^{\Ns} \fsi |\psisi(\r)|^2,
\end{equation}
where $\Ns$ is the number of occupied $\sigma$ spin orbitals.
The effective KS potential $V^{KS}_{\sigma}(\r)$ is composed of
external, Hartree, and exchange-correlation contributions,
\begin{equation}
\label{eq:VKS}
V^{KS}_{\sigma}(\r) = V_{ext}(\r) + V_H(\r) + \Vxcs^{KS}(\r),
\end{equation}
where the exchange-correlation potential $\Vxcs^{KS}(\r)$ is given by
\begin{equation}
\label{eq:VXC}
\Vxcs^{KS}(\r) = \frac{\delta \Exc[\{ n_{\uparrow}, n_{\downarrow} \}]}{\delta \ns(\r)}.
\end{equation}
If the exchange-correlation functional is explicitly orbital-dependent
$\Exc[\{ n_{\uparrow}, n_{\downarrow} \}] \equiv \Exc[\{ \psisi \}]$,
the corresponding $\Vxcs^{KS}(\r)=\Vxcs^{OEP}(\r)$ can be obtained by the
OEP integral equation\cite{Sha53,Tal76}
\begin{equation}
\label{eq:OEP}
\sum_{\si} \fsi \int d\r' [\Vxcs^{OEP}(\r') - \uxcsi(\r')] \psisi^\ast(\r')
	G_{\si}(\r',\r) \psisi(\r) \ + \ c.c. \ = \ 0,
\end{equation}
where the quantity $\uxcsi(\r)$ is given by
\begin{equation}
\uxcsi(\r)
	\equiv \frac{1}{\fsi \psisi(\r)}
	\frac{\delta \Exc[\{\psisi\}]}{\delta \psisi^\ast(\r)}.
\end{equation}

The KLI approximation to the OEP is obtained by ignoring certain
terms whose average over the ground-state spin density
vanishes.\cite{Kri90} It can be also heuristically derived by
approximating $(\epsilon_{\sj}-\epsilon_{\si})^{-1}$, appearing in
the Green's function in Eq. (\ref{eq:OEP}) by a constant $\Delta
\epsilon_{\sigma}:$\cite{Kri90}
\begin{equation}
\label{eq:Green}
G_{\si}(\r',\r)
	\equiv \sum_{j \neq i}  \frac{\psisj(\r') \psisj^\ast(\r)}
		{\epsilon_{\sj} - \epsilon_{\si}}
	\approx \frac{\sum_{j \neq i} \psisj(\r') \psisj^\ast(\r)}{\Delta \epsilon_\sigma},
\end{equation}
in which case Eq. (\ref{eq:Green}) reduces to integrals over the
wave functions independent of the spectrum. Slater's
approximation\cite{Sla51} can be obtained by simplifying the
numerator of Eq. (\ref{eq:Green}) further to $\sum_{j \neq i}
\psisj(\r') \psisj^\ast(\r) \approx \delta(\r-\r')$, thus ignoring
$-\psisi(\r') \psisi^\ast(\r)$ term.\cite{Sha53,Kri90} Although
the KLI approximation is generally applicable to any orbital
dependent exchange-correlation energy functional, we will
concentrate on the case of the exact exchange energy functional
defined as
\begin{equation}
\label{eq:EXX}
\Ex[\{ \psisi \}] =
-\frac{1}{2} \sum_{\sigma=\uparrow,\downarrow} \sum_{i,j}^{\Ns}
\fsi \fsj
\int d\r \int d\r'
\frac{\psisi^{*}(\r) \psisj^{*}(\r') \psisi(\r') \psisj(\r)}{|\r - \r'|},
\end{equation}
which has been actually used in this work. Here note that $\{ \psisi
\}$ are KS orbitals rather than HF orbitals. In this case,
$\uxsi(\r)$ is just the HF local exchange potential expression
evaluated with KS (instead of HF) orbitals $\{ \psisi \}$
\begin{equation}
\label{eq:UXSI}
\uxsi(\r) = - \frac{1}{\psisi(\r)}\sum_{j=1}^{\Ns} \fsj \psisj(\r) K_{\sji}(\r),
\end{equation}
where
\begin{equation}
\label{eq:KSJI}
K_{\sji}(\r) = \int d\r'\frac{\psisj^\ast(\r')\psisi(\r')}{|\r-\r'|}.
\end{equation}
The approximation of Eq. (\ref{eq:Green}) leads to the KLI-x potential
\begin{equation}
\label{eq:KLI}
\Vxs^{KLI-x}(\r) = \Vxs^{S}(\r) +
	\sum_{i=1}^{\Ns}\frac{\nsi(\r)}{\ns(\r)} (\VBxsi^{KLI-x}-\uBxsi),
\end{equation}
where
\begin{equation}
\Vxs^{S}(\r)
	= \sum_{i=1}^{\Ns} \frac{\nsi(\r)}{\ns(\r)}\uxsi(\r)
\end{equation}
is the Slater exchange potential,\cite{Sla51} with
$\nsi(\r)= \fsi |\psisi(\r)|^2$.
The quantities $\VBxsi^{KLI-x}$ and $\uBxsi$ are the expectation values of
$\Vx^{KLI-x}(\r)$ and $\uxsi(\r)$ with respect to the KLI-x orbital $\si$,
\begin{equation}
\VBxsi^{KLI-x} = \langle \psisi|\Vx^{KLI-x}|\psisi \rangle;
\ \uBxsi = \langle \psisi|\uxsi|\psisi \rangle.
\end{equation}
The condition
\begin{equation}
\label{eq:vx_HOMO}
\bar{V}_{x, \sigma \Ns}^{KLI-x}=\bar{u}_{x,\sigma \Ns}
\end{equation}
is automatically satisfied with the exact exchange energy functional,\cite{Gra98}
so only the remaining $\Ns-1$ constants $(\VBxsi^{KLI-x}-\uBxsi)$ in Eq. (\ref{eq:KLI})
need to be determined by the solution of the following linear equation:
\begin{equation}
\sum_{i=1}^{\Ns-1}(\delta_{ji}-M_{\sji}) (\VBxsi^{KLI-x}-\uBxsi)
= (\VBxsi^S-\uBxsi),
\end{equation}
where
\begin{equation}
M_{\sji} \equiv \int d\r \frac{\nsj(\r) \nsi(\r)}{\ns(\r)}, \
i,j = 1,\ldots,\Ns-1.
\end{equation}

The Slater potential $\Vxs^{S}(\r)$ in Eq. (\ref{eq:KLI}) ensures
the correct asymptotic behavior of $-1/r$ for large $r$ when the
exact exchange energy expression is employed.\cite{Tal76,Kri90} It
is known that $\Vxs^{S}(\r)$ alone is too deep,\cite{Kri90} and is
a factor of $3/2$ larger than $\Vxs^{KS}=\Vxs^{LSD}$ in the
homogeneous-electron-gas limit. The second term in Eq.
(\ref{eq:KLI}), which was originally missing in the Slater's
proposition, enables $\Vxs^{KLI-x}(\r)$ to correctly reduce to
$\Vxs^{LSD}$ in the homogeneous-electron-gas limit, and to
preserve the property of the integer discontinuity. \cite{Kri90}
Also, due to Eq. (\ref{eq:vx_HOMO}), the KLI-x highest occupied
eigenvalues satisfy the following condition
\begin{eqnarray}
\label{eq:ev_HOMO}
\epsilon_{\sigma\Ns}^{KLI-x} = \langle
\psi_{\sigma \Ns}|
	\bigl(-\frac{1}{2} \nabla^2 + V_{ext} + V_{H} + \Vx^{KLI-x} \bigr)
	|\psi_{\sigma \Ns} \rangle
\nonumber\\
= \langle \psi_{\sigma \Ns}|
	\bigl(-\frac{1}{2} \nabla^2 + V_{ext} + V_{H} + \bar{u}_{x,\sigma \Ns} \bigr)
	|\psi_{\sigma \Ns} \rangle
=\epsilon_{\sigma\Ns}^{HF'},
\end{eqnarray}
where $\epsilon_{\sigma\Ns}^{HF'}$ is the HF orbital energy
expression evaluated with the KS (rather than HF) orbital
$\psi_{\sigma\Ns}$.

At this point, we would like to emphasize that the KLI-x KS scheme
yields a local potential, while the HF method gives a nonlocal
exchange potential. It is well known that the HF method gives
eigenfunctions and eigenvalues for the unoccupied orbitals which
are inappropriate for describing actual excited states, because HF
unoccupied orbitals do not have self-interaction corrections.
\cite{Sla69} This point will be further emphasized and discussed
in Sec. \ref{subsec:eigenvalues} with specific examples.

\subsection{Computational Methods}
\label{subsec:method}

All the calculations in this work are based on the
finite-difference pseudopotential scheme\cite{Che94} which we have
previously applied to the study of neutral and charged water
clusters.\cite{YHK2} In this formulation, quantities such as KS
orbitals, densities, and potentials are expressed on grids in a
rectangular simulation box, and the Laplacians or gradients are
evaluated by higher-order finite-difference expressions. The most
computationally demanding step in the construction of the KLI-x
potential is the computation of the $(\Ns-1) \Ns/2$ pairs $\int
d\r'\psisj^\ast(\r')\psisi(\r')/|\r-\r'|$ in Eq.(\ref{eq:UXSI}).
Since these terms have the same structure as that of the Hartree
potential with density $\ns(\r')$ replaced by
$\psisj^\ast(\r')\psisi(\r')$, we use the techniques we have
developed for the solution of the Poisson equation, and solve the
discretized Poisson-like equations
\begin{equation}
\nabla^2 K_{\sji}(\r) = - 4 \pi \psisj^\ast(\r)\psisi(\r).
\end{equation}
To do so, we first generated boundary values for $K_{\sji}(\r)$ at
six sides of the simulation box using a multipole expansion.
Multipole expansions have been performed until the boundary value
error $|\delta K_{\sji}(\r_{boundary})|/|K_{\sji}(\r_{boundary})|$
was less than $10^{-3}$. Next, a coarse solution for $K_{\sji}(\r)$
on the entire simulation grid has been generated by the fast Fourier
transform method using a lower-order (3 points along each direction)
finite-difference expression. This solution has been subsequently
relaxed by the iterative preconditioned conjugate gradient method
using a higher-order (typically 13 points along each direction)
finite-difference method. Conjugate gradient relaxation steps have
been performed until the error $|\delta K_{\sji}(\r)|/|K_{\sji}(\r)|$
is decreased to below $10^{-3}$. For the preconditioning, a smoothing
type preconditioner has been used. We observed that potential mixing
rather than density mixing is more appropriate, which should be
understandable considering that the KLI-x potential is
orbital-dependent.

In addition, for the GGA calculations, we have constructed GGA
potentials with the scheme of White and Bird.\cite{Whi94} For other
computational details including the implementation of the GGA
potential, we refer to our other publications.
\cite{YHK2,Lee98,Lee97}

\subsection{Test calculations}
\label{subsec:test}

To assess the accuracy of our 3D KLI-x implementation, we considered
atoms and compared LDA and KLI-x results with those from an accurate
1D radial atomic code. One test was for an Appelbaum-Hamann type
local pseudopotential\cite{App73} modified to bind 5 electrons
$V(r)=-\frac{5}{r} erf(\sqrt{\alpha} r) + ( v_1 + v_2 r^2) e^{-\alpha
r^2}$, where $\alpha = 0.6102$, $v_1 = 3.042$, and $v_2  = -1.732$.
This choice was made to reconcile the limitations of the 3D code that
it can deal with only smooth pseudopotentials and the 1D code that it
can only produce accurate results for spherical atoms because of the
central field approximation. Using a grid spacing $h$ = 0.4 a.u.,
simulation cell length $L$ = 32 a.u., and finite-difference order 13,
we obtained agreement with the radial code of $\le 10^{-3}$ a.u./atom
for both LDA and KLI-x total energies. For the hydrogen atom, we have
used a local potential derived by Giannozzi\cite{Gyg93} and again
obtained an agreement of $\le 10^{-3}$ a.u./atom using $h$ = 0.4 a.u.
For several other atoms considered in the next section (C, N, and O),
we generated LDA pseudopotentials with the method of Troullier and
Martins \cite{Tro91} and compared LDA total energies and eigenvalues
from the two codes. With $h$ = 0.25 a.u., the 3D code reproduced the
1D atomic total energy with an accuracy of $\le 5 \times 10^{-3}$
a.u./atom which is sufficient for our purpose.

Before closing this section, we would like to comment on the issues
involved with the use of pseudopotentials. Although the rigorous and
consistent procedure would be to employ pseudopotentials generated
within the same functional approximations, we used LDA generated
pseudopotentials to carry out all other functional calculations for
two reasons. First, we believe this procedure will not change the
qualitative picture, since structural and electronic properties are
rather insensitive on the nature of the exchange-correlation energy
functional that is used for the small 1s cores of the first-row atoms
considered here.\cite{Woo88} Second, we note ambiguities associated
with pseudopotential generation procedures. For example, we observed
a small but sharp peak at the atomic center in the PBE GGA
pseudopoential, which caused a serious convergence
problem.\cite{Lee97} Moreover, in the KLI-x pseudopotential
generation, there is an intrinsic problem of slow decaying tails due
to the nonlocal nature of the exact exchange functional that somewhat
influences the results.\cite{Byl95,Sta98} These anomalous behaviors
can be only alleviated by some post-processings which introduce
additional arbitrariness in pseudopotenials. \cite{Byl95,Lee97}

%% ------------------------------------------------------------------- %%
\section{Results and Discussion}
\label{sec:results}

\subsection{Structural properties}
\label{subsec:structure}

%{\it Diatomic molecules}

We first examine the structural properties of $H_2$, $CO$, $N_2$,
$H_2O$, and $CH_4$ given in Table \ref{tab:diatom_ground} and Table
\ref{tab:polyatom_ground}. The binding energies $E_B$, equilibrium
bond lengths $r_e$, and vibrational frequencies $\omega_e$ of the
diatomic molecules were determined from the total energy versus
bond-length curves by fitting to the five parameter ($D_e$, $r_e$,
$\omega_e$, a, and b) Hulburt-Hirschfelder function,
\cite{Hul41,Hur76}
\begin{equation}
V(r) = D_e \ \bigl[(1-e^{-\beta x})^2 + b \beta^3 x^3 e^{-2\beta x}(1+ a \beta x) -1],
\end{equation}
where $x=r-r_e$ and $r_e$ and $D_e$ are the equilibrium distance and
bonding energy, using the simulated annealing method. \cite{NR} The
geometries of the polyatomic molecules have been optimized by
employing Hellmann-Feynman quantum forces. A grid spacing $h \ = \
0.25$ a.u. has been used for all the molecular calculations. For the
calculation of binding energies, the energies of each pseudoatom and
molecules have been calculated by the same method and with the same
$h$. We expect this procedure will result in a systematic
cancellation of errors, and in addition we may treat the atom without
central field approximation.

Our LDA and GGA results are essentially a reproduction of previous
studies. \cite{Pai82,Kut92,Pat97} The LDA consistently
overestimates the stability of the molecules and the GGA typically
cures this tendency of overestimation to a large extent. The HF
approximation on the other hand substantially underestimates the
binding energy. \cite{Hur76,Sza89} Bond lengths calculated by the
LDA are usually overestimated and one obtains corresponding
underestimations of vibrational frequencies for the diatomic
molecules. The HF method gives the opposite behavior. The KLI-x
approximation typically gives similar results as HF for the
structural properties which are mostly determined by the total
energy functional.\cite{Sza89} It is well known that the HF method
gives the wrong sign of the dipole moment of $CO$, and the LDA
corrects this. Again, the KLI-x calculation yields the wrong sign
of the dipole moment of $CO$ in agreement with the HF method. The
difference between magnitudes of the KLI-x and HF dipole moment of
$CO$ results from the fact that the HF value has been evaluated at
the experimental bond length. Our KLI-x calculation at the
experimental bond length gave $-0.275$ Debye.

The addition of an LDA or GGA correlation energy functional to the
exact exchange energy gives mixed results: Binding energies are
increased, hence improved the KLI-x values, but bond lengths are
decreased and become worse. So, we can conclude that the two
correlation functional approximations do not give systematic
improvements over the exact exchange energy functional, which
indicates that these approximations are not accurate descriptions
of correlation effects in molecules.

\subsection{KS Eigenvalues and Excitations}
\label{subsec:eigenvalues}

In Fig. \ref{fig:H2_vxc}, we compare the LDA-xc and KLI-x
potentials of $H_2$ to reemphasize the different nature of the
exchange(-correlation) potentials from the two methods. While the
LDA-xc potential decays exponentially, the KLI-x potential
reproduces the correct $-1/r$ asymptotic behavior, and this causes
marked differences in the eigenvalue spectra. This $-1/r$ decay
property lets the high-lying KS unoccupied eigenvalues correctly
approach a Rydberg series, and is also important for other
properties that are sensitive to the outer part of the charge
density, such as polarizabilities.

Eigenvalues of the highest occupied orbitals of the molecules
considered in the previous section at their equilibrium geometries
(Tables \ref{tab:diatom_ground} and \ref{tab:polyatom_ground}) are
listed in Table \ref{tab:ioniz_potential}. Compared with the LDA
or GGA, the KLI-x values agree very well with experimental
ionization potentials, and they are very close to HF values. It is
well known that HF gives a good approximation to the ionization
potential via Koopman's theorem due to the fact that omission of
correlation tends to be cancelled by the neglect of relaxation in
the ``frozen orbital'' approximation.\cite{Sza89} On the other
hand, in the exact DFT, the highest occupied orbital eigenvalue
equals the ionization potential without any relaxation
correction\cite{Alm85,Per82} as we typically see in our KLI-x
calculations, hence, in principle DFT should yield highest
occupied orbital eigenvalues in better agreement with experimental
ionization potentials than HF. However, in the usual LDA or GGA,
the highest occupied eigenvalues are in serious errors due to the
wrong decay property of exchange-correlation potential exemplified
in Fig. \ref{fig:H2_vxc}. The errors are about 0.2 a.u. for the
molecules we have studied as shown in Table
\ref{tab:ioniz_potential}. On the other hand, due to Eq.
(\ref{eq:ev_HOMO}), KLI-x should give highest occupied orbital
eigenvalues in good agreement with experimental ionization
potentials when HF values also agree with experimental ionization
potential. The addition of LDA or GGA correlation potentials to
the KLI-x potential usually lowers KLI-x highest occupied orbital
eigenvalues, but this leads to too large ionization potentials.

Among the molecules examined in this work, $N_2$ is a particularly
interesting case which shows the advantage and disadvantage of the
KLI-x scheme at the same time. First, note that the HF results are in
qualitative disagreement with experiment: it puts the $1\pi_u$ state
higher than $3\sigma_g$ state as the highest occupied orbital in
contrast to the experimental data. This incorrect ordering of the
first two ionization potentials is a well-known example of the
breakdown of the HF picture and indicates a large effect of
correlation upon the orbitals in $N_2$. \cite{Sza89} On the other
hand note that KLI-x gives the correct ordering of the two orbitals,
which implies that the local nature of the KLI-x potential can make a
difference even for the occupied-orbital eigenvalue spectrum. The
fact that the LDA also gives the correct ordering (although the value
is worse than the KLI-x one due to the above-mentioned reasons) shows
that there are cases where the local potential in the KS scheme is
superior to the nonlocal HF potential. The highest occupied state in
the KLI-x scheme $\epsilon_{3\sigma_g}^{KLI-x}$ is equal to
$\epsilon_{3\sigma_g}^{HF'}$ [according to Eq. (\ref{eq:ev_HOMO})]
which results in the biggest KLI-x error among the molecules studied
since this also has the largest HF error. Better quantitative
agreement with experiment can be obtained by incorporating the
correlations, e.g. through many-body perturbation type
approach.\cite{Sza89}

%-- UNOCCUPIED ORBITAL EIGENVALUES --

In addition to the eigenvalues of the highest occupied orbitals, we
have also examined eigenvalue differences between the highest
occupied and the lowest unoccupied orbitals, which are compared with
experimental lowest triplet and singlet vertical excitation energies
in Table \ref{tab:exctn_energy}. First, we point out that the HF
calculations give positive orbital energies for all the virtual
orbitals in the molecules considered here.\cite{Hur76,Sza89} This is
because unoccupied orbitals in HF do not really correspond to excited
states of the system, in which the excited state electron would have
been removed from one of the lower states and acted on by $N-1$
remaining electrons,\cite{Sla69} but rather the bound states (if any)
of the singly-charged negative ion, in which the extra electron sees
the field due to other $N$ electrons.\cite{Hur76} To make this
argument more explicit, we follow Slater\cite{Sla63} and rewrite Eqs.
(\ref{eq:UXSI}) and (\ref{eq:KSJI}) (now using the HF orbitals
$\psisi^{HF}$ instead of KS orbitals $\psisi$, so $\uxsi^{HF}$ is the
true HF local exchange potential) as
\begin{equation}
 \uxsi^{HF}(\r) =
 - \int d\r'
 \frac{
 \bigl[ \sum_{j=1}^{\Ns} \fsj
 \psisi^{HF \ast}(\r) \psisj^{HF \ast}(\r') \psisj^{HF}(\r) \psisi^{HF}(\r')
 \bigr]
 / \bigr[ \psisi^{HF \ast}(\r) \psisi^{HF}(\r) \bigr] }
 {|\r-\r'|},
\end{equation}
and identify
\begin{equation}
\label{eq:RHOX}
 -\frac{\sum_{j=1}^{\Ns} \fsj
 \psisi^{HF \ast}(\r) \psisj^{HF \ast}(\r') \psisj^{HF}(\r) \psisi^{HF}(\r')}
 {\psisi^{HF \ast}(\r) \psisi^{HF}(\r)}
\end{equation}
as an exchange charge density. Because the $\psisi^{HF}$'s are
orthonormal, the exchange charge density integrated over $d\r'$ will
be minus one for the occupied orbital ($j=i$ term exists in the
summation $\sum_{j=1}^{\Ns}$ and $\fsi=1$), but zero for the
unoccupied orbital ($\psisi^{HF}$ is not included in
$\sum_{j=1}^{\Ns}$). Hence, the self-consistent field in which an
unoccupied orbital moves is that of the nucleus and all $N$ electrons
instead of $N-1$ electrons.

On the other hand, in DFT, as mentioned in Sec. \ref{sec:intro},
differences of KS eigenvalues are well-defined approximations for
excitation energies. \cite{Gor96} Umrigar {\it et al.} have confirmed
that this is the case for several atomic systems by showing that the
exact KS eigenvalue differences obtained from KS potentials derived
from accurate QMC densities almost always lie between the triplet and
singlet experimental excitation energies.\cite{Umr98} They further
claimed that this is because of the fact that, in addition to the
well known exact properties of $\Vxc^{KS}$ (Sec. \ref{sec:intro}),
$\Vxc^{KS}$  agrees with the quasiparticle amplitude\cite{Alm85} for
$r \rightarrow \infty$ not only up to $1/r$ but up to order $1/r^4$
inclusive. However, in the LDA or GGA, no bound unoccupied state
exists in many cases, due to the shallow nature of the corresponding
exchange-correlation potentials. We can expect that the KLI-x is a
much better approximation than the LDA/GGA, since it satisfies Eq.
(\ref{eq:ev_HOMO}) and has the correct asymptotic $-1/r$ behavior.
Our KLI-x results indeed always give negative virtual orbital
eigenvalues for the studied molecules and also show good agreement
with experimental values. The single exception is $H_2O$ in which the
difference between the highest occupied and lowest unoccupied orbital
eigenvalues is bigger than the experimental triplet and singlet
excitation energies by $\approx 0.07$ a.u. Further detailed
agreements with different multiplet states can be obtained by
employing a more involved theory such as DFT perturbation
theory\cite{Gor96,Fil97} or time-dependent DFT.\cite{Gro96,Cas96}

We close this section by reconsidering the meaning of eigenvalues in
the OEP-x (or KLI-x) and HF schemes to make the differences between
the two be clear. In the fractional occupation number formalism, KLI
have shown that
\begin{equation}
\label{eq:Janak}
 \frac{\partial E}{\partial f_i}
	= \langle \psisi| (-\frac{1}{2} \nabla^2 + V_{ext} + V_{H} + \uxsi ) |\psisi \rangle
	\equiv \epsilon_{\si}^{HF'},
\end{equation}
where $\epsilon_{\si}^{HF'}$ is the HF orbital energy expression for
the orbital $\si$ evaluated using the OEP-x orbitals $\psisi$.
\cite{Kri90} We note that this can be generalized to the correlated
OEP case. \cite{Kri90,Cas99} Equation (\ref{eq:Janak}) can be
rearranged to the form in which the relationship between the two
methods is transparent:
\begin{equation}
\label{eq:OEP_HF}
 \epsilon_{\si}^{OEP-x} + \langle \psisi| (\uxsi -
 \Vx^{OEP-x}) |\psisi \rangle = \epsilon_{\si}^{HF'}.
\end{equation}
Note that from the condition $\bar{V}_{x,\sigma\Ns}^{OEP-x}=\bar{u}_{x,\sigma\Ns}$,
satisfied by both OEP-x and KLI-x [Eq. (\ref{eq:vx_HOMO})],
Eq. (\ref{eq:ev_HOMO}) is reproduced from Eq. (\ref{eq:OEP_HF}) for the $\sigma \Ns$ state.
Hence, the eigenvalue difference between the highest occupied orbital state $\psi_{\sigma\Ns}$
and the unoccupied orbital state $\psi_{\sigma a}$ is given by
\begin{eqnarray}
 \epsilon_{\sigma a}^{OEP-x} - \epsilon_{\sigma \Ns}^{OEP-x}
+ \langle \psi_{\sigma a}| (u_{x,\sigma a} - \Vx^{OEP-x}) |\psi_{\sigma a} \rangle \nonumber\\
= \epsilon_{\sigma a}^{HF'} - \epsilon_{\sigma \Ns}^{HF'}.
\end{eqnarray}
Assuming that the differences between the HF and OEP-x orbitals are
negligible, we can see that HF orbital eigenvalue differences are
much bigger than OEP-x values, because the $\langle \psi_{\sigma a}|
u_{x,\sigma a} |\psi_{\sigma a} \rangle$ term is small due to the
property of the exchange charge density [Eq. (\ref{eq:RHOX})] for the
unoccupied orbital as discussed above, while the $\langle
\psi_{\sigma a}| \Vx^{OEP-x} |\psi_{\sigma a} \rangle$ term is
clearly quite large.

%% ------------------------------------------------------------------- %%
\section{Conclusions}
\label{sec:conclusions}

In this work, we have studied structural and excitation properties of
small molecules using the KLI approximation to the OEP with the exact
exchange energy functional only (KLI-x) and augmented by LDA and GGA
correlation functionals. For structural properties, the KLI-x gave
comparable results to those of the HF method. For excitation
properties, the KLI-x results in good eigenvalue spectra for {\it
both the highest occupied and unoccupied} orbitals, since it has the
correct asymptotic large $r$ behavior of $\Vxc^{KS}$ unlike the LDA
or GGA. Especially, we find that unoccupied orbital eigenvalues are
better described by KLI-x than by HF, which illustrates the advantage
of the KS scheme in general.

%% ------------------------------------------------------------------- %%
\acknowledgments

This work has been supported by the National Science Foundation under
Grant No. DMR 98-0273 (Y.-H. K. and R. M. M.) and by the Office of
Naval Research under Grant No. N0014-98-1-0594 (M. S.). We
acknowledge Profs. J. B. Krieger and J. P. Perdew for critical
readings, Prof. E. K. U. Gross for helpful discussions, Prof. J.
Soler for providing us with his GGA exchange-correlation routine, and
Dr. D. Sanchez-Portal for performing GGA calculations for
comparisons.

%% ------------------------------------------------------------------- %%

%% ------------------------------------------------------------------- %%
\newpage

\begin{table}
\caption{Binding energies, bond lengths, and vibrational frequencies
of $H_2$, $CO$, and $N_2$, calculated with different methods.
In addition, the dipole moment of $CO$ has been given,
whose sign has been defined such that $C^{-}O^{+}$ is positive.
Energies and bond lengths are given in a.u., vibrational frequencies
in $cm^{-1}$, and dipole moments in Debye.}
\begin{tabular}{ldddddddddd}
 &\multicolumn{3}{c}{$H_2$}  &\multicolumn{4}{c}{$CO$}
 &\multicolumn{3}{c}{$N_2$}\\
 &$E_B$ &$r_e$ &$\omega_e$ &$E_B$ &$r_e$ &$\omega_e$  &$\mu$
 &$E_B$ &$r_e$ &$\omega_e$ \\
 \hline
 LDA-xc &0.184 &1.44 &4228  &0.466 &2.13 &2193 &0.248   &0.396 &2.09 &2385 \\
 GGA-xc &0.168 &1.41 &4228  &0.424 &2.12 &2166 &0.265   &0.368 &2.09 &2383 \\
 \hline
 KLI-x  &0.136 &1.39 &4647  &0.287 &2.06 &2445 &-0.163  &0.176 &2.01 &2634 \\
 +LDA-c &0.185 &1.36 &4765  &0.342 &2.05 &2595 &-0.184  &0.271 &2.00 &2710 \\
 +GGA-c &0.166 &1.38 &4734  &0.361 &2.06 &2641 &-0.170  &0.273 &2.00 &2805 \\
 \hline
 HF-x\tablenotemark[1]
		&0.134 &1.39 &4582  &0.293 &2.08 &2431 &-0.279\tablenotemark[2]
														&0.195 &2.01 &2730   \\
 \hline
 Exp.\tablenotemark[1]
		&0.174 &1.40 &4400  &0.414 &2.13 &2170 &0.123   &0.364 &2.07 &2358   \\
\end{tabular}
\tablenotetext[1]{Reference \cite{Hur76}.}
\tablenotetext[2]{At the experimental bond length.}
\label{tab:diatom_ground}
\end{table}

\begin{table}
\caption{Binding energies and bond lengths of $CH_4$ and $H_2O$, and
the bond angle of $H_2O$. Binding energies and bond lengths are given
in a.u., bond angles are given in degrees.}
\begin{tabular}{lddddd}
	&\multicolumn{2}{c}{$CH_4$} & \multicolumn{3}{c}{$H_2O$} \\
	&$E_b$ &$r_e \ (C-H)$ &$E_b$  &$r_e \ (O-H)$  &$\theta \ (H-O-H)$ \\
 \hline
 LDA-xc &0.743 &2.07 &0.420 &1.82 &104.2 \\
 GGA-xc &0.681 &2.06 &0.370 &1.81 &104.5 \\
 \hline
 KLI-x  &0.536 &2.05 &0.257 &1.76 &106.0 \\
 +LDA-c &0.678 &2.04 &0.344 &1.76 &106.0 \\
 +GGA-c &0.672 &2.03 &0.339 &1.77 &105.3 \\
 \hline
 HF-x\tablenotemark[1]
		&0.522 &2.05 &0.258 &1.78 &106.1 \\
 \hline
 Exp.\tablenotemark[1]
		&0.668 &2.05 &0.374 &1.81 &104.5 \\
\end{tabular}
\tablenotetext[1]{Reference \cite{Hur76}.}
\label{tab:polyatom_ground}
\end{table}

\begin{table}
\caption{Absolute values of highest occupied orbital eigenvalues of
the $H_2$, $CO$, $N_2$, $H_2O$, and $CH_4$ at their equilibrium
geometries (Table I and II). For $N_2$, absolute values of the next
highest occupied orbital eigenvalues are also given. Note the
differet ordering of HF and KLI-x for this case. Experimental values
are ionization potentials. Energies are in a.u.}
\begin{tabular}{lddddd}
	&$H_2 \ (-\epsilon_{1\sigma_g})$
	&$CO \ (-\epsilon_{5\sigma})$
	&$N_2 \ (-\epsilon_{3\sigma_g}/-\epsilon_{1\pi_u})$
	&$H_2O \ (-\epsilon_{1b_1})$
	&$CH_4 \ (-\epsilon_{1t_2})$\\
\tableline
LDA-xc  &0.37   &0.34   &0.38/0.44  &0.26   &0.34   \\
GGA-xc  &0.38   &0.34   &0.38/0.43  &0.27   &0.34   \\
\hline
KLI-x   &0.60   &0.55   &0.64/0.69  &0.50   &0.54   \\
KLI-x + LDA-c
	&0.60   &0.61   &0.69/0.75  &0.56   &0.60   \\
KLI-x + GGA-c
	&0.58   &0.59   &0.67/0.74  &0.55   &0.58   \\
\hline
HF-x\tablenotemark[1]
	&0.60   &0.55   &0.64/0.62  &0.51   &0.55   \\
\hline
Exp.\tablenotemark[1]
	&0.58   &0.58   &0.57/0.62  &0.46   &0.53   \\
\end{tabular}
\tablenotetext[1]{Reference \cite{Hur76}.}
\label{tab:ioniz_potential}
\end{table}

\begin{table}
\caption{Differences between highest occupied and lowest unoccupied orbital
states eigenvalues of $H_2$, $CO$, $N_2$, $H_2O$, and $CH_4$
at their respective equilibrium geometries given in Table I and II.
A $-$ mark indicates that the lowest unoccupied orbital state is in the continuum
and hence unstable.
Experimental values are vertical excitation energies to final triplet and
singlet states shown in parentheses.
Energies are in a.u.}
\begin{tabular}{lccccc}
	&$H_2 \ (\epsilon_{1\sigma_u}-\epsilon_{1\sigma_g})$
	&$CO  \ (\epsilon_{2\pi}-\epsilon_{5\sigma})$
	&$N_2 \ (\epsilon_{1\pi_g}-\epsilon_{3\sigma_g})$
	&$H_2O \ (\epsilon_{4a_1}-\epsilon_{1b_1})$
	&$CH_4 \ (\epsilon_{3a_1}-\epsilon_{1t_2})$\\
\tableline
LDA-xc  &$-$    &0.26   &0.30   &$-$    &$-$    \\
GGA-xc  &$-$    &0.26   &0.31   &$-$    &$-$    \\
\hline
KLI-x   &0.47   &0.28   &0.34   &0.34   &0.42   \\
KLI-x + LDA-c
	&0.50   &0.28   &0.34   &0.36   &0.46   \\
KLI-x + GGA-c
	&0.49   &0.27   &0.34   &0.37   &0.45   \\
\hline
HF-x    &$-$    &$-$    &$-$    &$-$    &$-$    \\
\hline
Exp.
\tablenotemark[1]
	&0.42 ($^3\Sigma_u^+$)
	&0.23 ($^3\Pi$)
	&0.28 ($^3\Sigma_u^+$)
	&0.26 ($^3B_1$)
	&0.40 ($^3T_2$)
	\\
	&0.46 ($^1\Pi_u$)
	&0.31 ($^1\Pi$)
	&0.34 ($^1\Pi_g$)
	&0.28 ($^1B_1$)
	&0.41 ($^1T_2$)
\end{tabular}
\tablenotetext[1]{Reference \cite{Hur76}.}
\label{tab:exctn_energy}
\end{table}

%% ------------------------------------------------------------------- %%
\newpage

\begin{figure}
\caption{Comparison of the LDA exchange-correlation potential and the
KLI exchange potential for $H_2$ with $H$ atoms located at $\pm 0.7$
a.u. Note that the LDA exchange-correlation potential incorrectly
decays exponentially, while the KLI exchange potential decays with
$-1/r$ as in the exact Kohn-Sham exchange-correlation potential.}
\label{fig:H2_vxc}
\end{figure}

%% ------------------------------------------------------------------- %%
\newpage

  \begin{minipage}[H]{0.70\linewidth}
  \vspace{0.5cm}
  \centering\epsfig{file=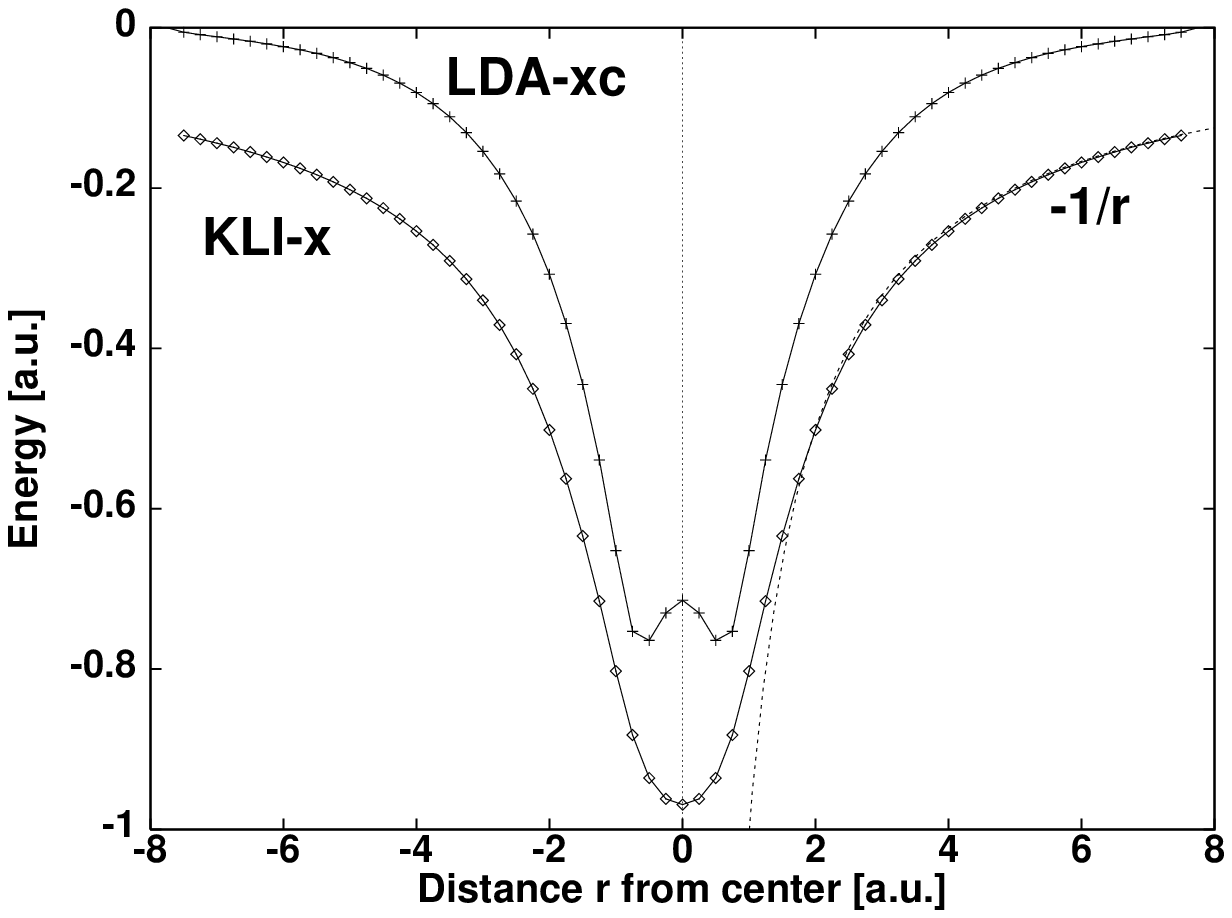,width=\linewidth}
  \end{minipage}  \hspace{1.2cm}
   Fig. 1

\end{document}